\documentstyle[12pt,worldsci]{article}
\def\be{\begin{equation}}
\def\ee{\end{equation}}
\def\bea{\begin{eqnarray}}
\def\eea{\end{eqnarray}}
\def\bq{\begin{quote}}
\def\eq{\end{quote}}
\def\bseq{\begin{subequation}}
\def\eseq{\end{subequation}}
\def\bsea{\begin{subeqnarray}}
\def\esea{\end{subeqnarray}}
\def\simlt{\mathrel{\lower2.5pt\vbox{\lineskip=0pt\baselineskip=0pt
           \hbox{$<$}\hbox{$\sim$}}}}
\def\simgt{\mathrel{\lower2.5pt\vbox{\lineskip=0pt\baselineskip=0pt
           \hbox{$>$}\hbox{$\sim$}}}}
\def\ov{\overline}
\def\ie{i.e.~}

\def\epem{e^+e^-}

\def\tb{\tan\beta}

\def\matb{(m_A,\tan\beta)}
\def\mpl{M_{\rm P}}

\def\msq{m_{\tilde{q}}}

\def\tev{\rm \; TeV}
\def\gev{\rm \; GeV}

\def\pb{\rm \; pb}
\def\sabsq{\sin^2 (\beta - \alpha)}
\def\cabsq{\cos^2 (\beta - \alpha)}
\begin{document}
\begin{titlepage}
\vspace*{-1cm}
\noindent
\phantom{bla}
\hfill{CERN-TH.6951/93}
\\
\phantom{DRAFT}
\hfill{hep-ph/9307293}
\vskip 5.0cm
\begin{center}
{\large Fabio Zwirner}\footnote{On leave from INFN, Sezione di
Padova, Padua,
Italy.}
\end{center}
\begin{center}
{\large
Theory Division, CERN \\
CH-1211 GENEVE 23 \\ Switzerland }
\end{center}
\vskip 3.0cm
\begin{center}
{\Large\bf THE SUSY WORLD}
\end{center}
\vskip 3.0cm
\begin{center}
{\em
Summary Talk of the Workshop `Properties of SUSY particles'
\\
Erice, September 28 -- October 4, 1992.}
\end{center}
\vfill{
CERN-TH.6951/93
\newline
\noindent
July 1993}
\end{titlepage}
\setcounter{footnote}{0}
As Fayet reminded us in his introductory talk$^{\ref{fayet}}$,
realistic models
of low-energy supersymmetry have been studied for more than 15 years.
At first
sight, the absence of direct experimental evidence does not match
such an
intense theoretical effort, and puts supersymmetry on the same
footing as many
other extensions of the Standard Model (SM). A number of theoretical
and
phenomenological reasons, however, make low-energy supersymmetry
particularly
attractive with respect to its alternatives: the intense activity
reported at
this workshop is there to prove it! My attempt to summarize its
highlights will
be organized as follows. In Section 1, I shall review the main
motivations that
lead most of us to consider the `SUSY world' as a plausible scenario.
The
simplest realization of low-energy supersymmetry, the Minimal
Supersymmetric
extension of the Standard Model (MSSM), will be recalled, with some
comments on
possible non-minimal variations. Section 2 will summarize the
phenomenology of
supersymmetric particle searches, including Higgs bosons, at present
and future
accelerators. Finally, Section 3 will review some open theoretical
problems
connected with spontaneous supersymmetry breaking in supergravity and
superstring models, and draw some conclusions. Unavoidably, the
selection of
topics will depend on space limitations and on my personal view of
the subject:
I apologize in advance with the participants whose contributions
would have
deserved a better treatment.
\vskip 1.0cm

\section{The MSSM: a paradigm for low-energy SUSY}

\subsection{Theoretical motivations}

As discussed in the talks by Fayet$^{\ref{fayet}}$ and
Kounnas$^{\ref{kounnas}}$, there are many good reasons to believe
that
four-dimensional supersymmetry and its local version, supergravity,
could be
relevant in a fundamental theory of particle interactions. In
particular,
superstrings are the present best candidates for a consistent quantum
theory
unifying gravity with all the other fundamental interactions, and
supersymmetry
seems to play a very important role for the quantum stability of
superstring
solutions in flat four-dimensional space-time. Experimental data,
however, tell
us that supersymmetry is broken, but strings have not yet given us
any insight
about the scale of supersymmetry breaking.

The only motivation for low-energy supersymmetry, \ie  supersymmetry
effectively broken around the electroweak scale, comes from the
{\it naturalness} or {\it hierarchy} problem of the SM, whose
formulation
will now be sketched. Despite its remarkable
phenomenological success, it is impossible not to regard the SM as an
effective low-energy theory, valid up to some energy scale $\Lambda$
at
which it is replaced by some more fundamental theory.
Certainly $\Lambda$ is less than the Planck scale $\mpl \sim 10^{19}
\gev$, since one needs a theory of quantum gravity to describe
physics
at these energies. However, the study of the Higgs sector of the SM
suggests that $\Lambda$ should rather be close to the Fermi scale,
$G_{\rm F}^{-1/2} \sim 300 \gev$. The argument goes as follows.
Consistency of
the SM requires the SM Higgs mass to be less than $O(1 \tev)$. If one
then tries to extend the validity of the SM to energy scales $\Lambda
\gg G_{\rm F}^{-1/2}$, one is faced with the fact that in the SM
there is
no symmetry to justify the smallness of the Higgs mass with respect
to the (physical) cut-off $\Lambda$. This is related to the existence
of
quadratically divergent loop corrections to the Higgs mass in the SM.
Motivated by this problem, much theoretical effort has been devoted
to finding descriptions of electroweak symmetry breaking, which
modify the
SM at scales $\Lambda \sim G_{\rm F}^{-1/2}$. Here supersymmetry
comes into
play because of its improved ultraviolet behaviour with respect to
ordinary
quantum field theories, due to cancellations between bosonic and
fermionic
loop diagrams. If one wants to have a low-energy effective Lagrangian
valid up
to scales $\Lambda \gg G_{\rm F}^{-1/2}$, with one or more elementary
scalar
fields, kept light without unnatural fine-tuning of parameters, the
solution is to introduce supersymmetry, effectively broken in the
vicinity
of the electroweak scale. This does not yet explain why the scale
$M_{\rm SUSY}$ of supersymmetry breaking is much smaller than
$\Lambda$
(further considerations on this problem will be made in the final
section),
but at least links the Fermi scale $G_{\rm F}^{-1/2}$ to the
supersymmetry-breaking scale $M_{\rm SUSY}$, and makes the hierarchy
$G_{\rm F}^{-1/2} \sim M_{\rm SUSY} << \Lambda$ stable against
radiative
corrections.

\subsection{The MSSM}

The most economical realization of low-energy supersymmetry is the
MSSM, whose
defining assumptions were recalled by Fayet$^{1}$. The gauge group is
$G=SU(3)_C \times SU(2)_L \times U(1)_Y$, and the matter content
corresponds to three generations of quarks and leptons, as in the SM.
To give masses to all charged fermions and to avoid chiral anomalies,
however, one is forced to introduce two complex Higgs doublets, one
more than in the SM case. To enforce baryon and lepton number
conservation
in renormalizable interactions, one imposes a discrete $R$-parity: in
practice, $R=+1$ for all ordinary particles (quarks, leptons,
gauge and Higgs bosons), $R = -1$ for their superpartners
(spin-0 squarks and sleptons, spin-$1/2$ gauginos and higgsinos).
A globally supersymmetric Lagrangian ${\cal L}_{\rm SUSY}$ is then
fully
determined by the superpotential (in standard notation)
\be
f = h^U Q U^c H_2 + h^D Q D^c H_1
+ h^E L E^c H_1 + \mu H_1 H_2 \, .
\ee

To proceed towards a realistic model, one has to introduce
supersymmetry breaking. In the absence of a fully satisfactory
mechanism for spontaneous supersymmetry breaking at a fundamental
level, it seems sensible to {\em parametrize} supersymmetry breaking
at
low energy by a collection of {\em soft} terms, ${\cal L}_{soft}$,
which preserve the good ultraviolet properties of global
supersymmetry.
This ${\cal L }_{soft}$ contains mass terms for scalar fields and
gauginos,
as well as a restricted set of scalar interaction terms
\be
\label{lsoft}
\begin{array}{ccl}
- {\cal L }_{soft}
& = &
\sum_i \tilde{m}_i^2 | \varphi_i|^2 + {1 \over 2} \sum_A M_A
\ov{\lambda}_A \lambda_A + \left( h^U A^U Q U^c H_2
\right.
\\
& + &
\left.
h^D A^D Q D^c H_1 + h^E A^E L E^c H_1 + m_3^2 H_1 H_2 + {\rm h.c.}
\right),
\end{array}
\ee
where $\varphi_i$ ($i=H_1,H_2,Q,U^c,D^c,L,E^c$) denotes the generic
spin-0
field, and $\lambda_A$ ($A=1,2,3$) the generic gaugino field. Observe
that,
since $A^U,A^D$ and $A^E$ are matrices in generation space, ${\cal L
}_{soft}$
contains in principle a huge number of free
parameters. Moreover, for generic values of these parameters one
encounters phenomenological problems with flavour-changing neutral
currents, new sources of CP violation, and charge- and
colour-breaking
vacua. All the above problems can be solved at once if one assumes
that the running mass parameters in ${\cal L}_{soft}$, defined at the
one-loop level and in a mass-independent renormalization scheme,
can be parametrized, at some grand-unification scale $M_U$, by a
universal gaugino mass $m_{1/2}$, a universal scalar mass $m_0$, and
a
universal trilinear scalar coupling $A$, whereas $m_3^2$ remains in
general
an independent parameter.

\subsection{Non-minimal models}

The above assumptions, which define the MSSM, are plausible but not
compulsory:
relaxing them leads to non-minimal supersymmetric extensions of the
SM.

For example, as discussed in the talks by Dreiner$^{\ref{dreiner}}$
and
Kobayashi$^{\ref{kobayashi}}$, one can consider models in which
$R$-parity is
explicitly broken by some superpotential couplings. The acceptable
ones have
{\em either} the baryon {\em or} the lepton number violated by
renormalizable
interactions among light particles, and give rise to phenomenological
signatures that can be drastically different from the ones of the
MSSM. A proof
of this is the fact that, with some luck, one could be able to detect
signals
of supersymmetry even at HERA, which in the case of the MSSM cannot
add much to
what we already know.

Another possibility is to enlarge the Higgs sector of the model, for
example
by adding a gauge-singlet Higgs superfield, as discussed in the talk
by
Kane$^{\ref{kane}}$. In this case the restrictions imposed by
supersymmetry on
Higgs masses and couplings are much weaker than in the minimal case.
On the
other hand, requiring perturbative unification of couplings can still
lead to
interesting constraints, and in particular to an upper bound on the
lightest
Higgs mass of the order of 150 GeV.

As for the boundary conditions at the unification scale, one can
observe with
Ib\'a\~nez$^{\ref{ibanez}}$ that the simplest unification conditions
on the
gauge coupling constants are not really compulsory in string
unification. In a
general four-dimensional string model with gauge group $SU(3) \times
SU(2)
\times U(1) \times G$, one can have tree-level relations such as $g_1
k_1 = g_2
k_2 = g_3 k_3$, where the integer numbers $k_a$ ($a=1,2,3$) are the
so-called
Kac-Moody levels. In  string unification there is no fundamental
reason for the
tree-level prediction $\sin^2 \theta_W \equiv (3/5) g_1^2 / [(3/5)
g_1^2 +
g_2^2 ] = 3/8$, which is so successful when combined with the MSSM
quantum
corrections. Such an occurrence could be related to the existence of
a gauge
$U(1)_X$ symmetry of the Peccei-Quinn type, whose anomaly is
cancelled by a
Green-Schwarz mechanism, but no realistic string model with these
properties
has been found yet.

Less radically, one can also consider the possibility of
non-universal boundary
conditions on the soft supersymmetry-breaking terms. Such a
possibility, which
could be realized in string model-building (an example was given in
the talk by
Antoniadis$^{\ref{antoniadis}}$, others were recently discussed in
Ref.~\ref{nonuniv}), is strongly constrained by the phenomenological
limits on
flavour-changing neutral currents, but would lead to modified
relations among
the low-energy parameters with respect to the MSSM case.

All these non-minimal extensions remind us that we should not take
the MSSM as
the only viable paradigm for low-energy supersymmetry, and that
experimental
analyses should rather rely on the smallest possible amount of
theoretical
assumptions. On the other hand, non-minimal models typically increase
the
number of free parameters without correspondingly increasing the
physical
motivation, so we shall not discuss them further.

\subsection{Phenomenological virtues of the MSSM}

It is perhaps useful, at this point in the discussion, to recall some
phenomenological virtues of the MSSM (besides the solution of the
`technical' part of the hierarchy problem), which were mentioned at
this
workshop.

As stressed in the talk by Haber$^{\ref{haber}}$, an aspect that
became
particularly relevant after the recent precision measurements at LEP
is the
fact that electroweak data put little indirect constraints, via
radiative
corrections, on the MSSM  parameters. In most of its parameter space,
the MSSM
predictions for electroweak observables coincide in practice with
those of the
SM for a relatively light Higgs. Deviations comparable to the present
experimental accuracy can only occur for a light stop-sbottom sector
with large
mass splittings, or for a chargino with mass just above the
production
threshold at LEP~I. This is not the case, for example, of technicolor
and
extended technicolor models, which are severely constrained by the
recent LEP
data.

Another important property of the MSSM, discussed in the talks by
Haber$^{\ref{haber}}$ and Wagner$^{\ref{wagner}}$, is related to the
fact that
the running top Yukawa coupling $h_t(Q)$ has an effective infrared
fixed point,
smaller than in the SM case. Neglecting mixing and the Yukawa
couplings of the
remaining fermions, $h_t$ obeys the following one-loop
renormalization group
equation (RGE)
\be
\label{yukawa}
\frac{d h_t}{d t} =  \frac{h_t}{8 \pi^2}
\left(
- \frac{8}{3} g_3^2
- \frac{3}{2} g_2^2
- \frac{13}{18} g'^2
+ 3 h_t^2
\right) \, ,
\;\;\;\;\;
(t \equiv \log Q) \, .
\ee
A close look at Eq.~(\ref{yukawa}) can give us some important
information
about the acceptable values for the top-quark mass, $m_t$, and the
ratio of
vacuum expectation values, $\tb \equiv v_2 / v_1$.  However high is
the value
assigned to it at the unification scale, $h_t$ evaluated at the
electroweak
scale never exceeds a certain maximum value $h_t^{max} \simeq 1$.
This implies
that, for any given value of $\tb$, there is a corresponding maximum
value for
the top-quark mass. A na\"{\i}ve one-loop calculation gives
$m_t^{max} \sim 200
\gev \cdot \sin \beta$, which naturally puts the top-quark mass in
the range
presently allowed by direct searches and electroweak precision data.
The
results of a more refined calculation, which includes the effects of
all
third-generation Yukawa couplings and of the supersymmetric threshold
$M_{\rm
SUSY}$, are shown in Fig.~1.
\begin{figure}
\vspace{9.0cm}
\caption{The region of the $(\tb,m_t)$ parameter space in which all
running
Yukawa couplings remain finite at energy scales up to $\Lambda =
10^{16}
{\rm \; GeV}$ (from Ref.~9).}
\end{figure}

In the MSSM, $R$-parity makes the lightest $R$-odd supersymmetric
particle
(LSP) absolutely stable. In most of the otherwise acceptable
parameter space,
the LSP is neutral and weakly interacting, rarely a sneutrino, and
typically
the lightest, $\tilde{\chi}$, of the neutralinos (the mass
eigenstates of
the neutral gaugino-higgsino sector). Then the LSP is a natural
candidate
for cold dark matter, as discussed by
Roszkowski$^{\ref{roszkowski}}$, who
reported calculations of the neutralino relic density in different
regions of
the MSSM parameter space. His results could be summarized as follows.
In most
of the otherwise acceptable parameter space, the LSP is
cosmologically
harmless, in the sense that its relic density is smaller than the
closure
density of the Universe. Moreover, in a small but non-negligible
region of
parameter space, the LSP relic density turns out to be large enough
to be of
cosmological interest in relation with the dark-matter problem. It is
then
conceivable, even if not very likely, that the first evidence for
supersymmetry
could come from the dedicated experiments searching for dark-matter
signals!

One of the most attractive features of the MSSM is the possibility
of describing the spontaneous breaking of the electroweak gauge
symmetry as an effect of radiative corrections, as discussed by
Kounnas$^{\ref{kounnas}}$ and Lahanas$^{\ref{lahanas}}$. It is
remarkable
that, starting from universal boundary conditions at the unification
scale, it is possible to explain naturally why fields carrying colour
or electric charge do not acquire non-vanishing VEVs, whereas the
neutral
components of the Higgs doublets do. We give here a simplified
description
of the mechanism in which the physical content is transparent. The
starting
point is a set of boundary conditions on the independent model
parameters
at the unification scale $Q = M_U$. One then evolves all the running
parameters from the grand-unification scale to a low scale $Q \sim
G_F^{-1/2}$,
according to the RGEs, and considers the
renormalization-group-improved
tree-level potential
\be
\label{v0}
\begin{array}{ccl}
V_0 (Q) & = &
m_1^2 \left| H_1 \right|^2 +
m_2^2 \left| H_2 \right|^2 +
m_3^2 \left( H_1 H_2 + {\rm h.c.} \right)
\\ & & \\ & + &
\displaystyle{
{g^2 \over 8} \left( H_2^{\dagger} {\vec\sigma} H_2
+ H_1^{\dagger} {\vec\sigma} H_1 \right)^2 +
{g'^{\, 2} \over 8} \left(  \left| H_2 \right|^2 -
\left| H_1 \right|^2 \right)^2} \, .
\end{array}
\ee
All masses and coupling constants in $V_0(Q)$ are running parameters,
evaluated at the scale $Q$. The minimization of the potential in
Eq.~(\ref{v0}) is straightforward. To generate non-vanishing VEVs
$v_1 \equiv
\langle H_1^0 \rangle$ and $v_2 \equiv \langle H_2^0 \rangle$, one
needs
\be
{\cal B} \equiv m_1^2 m_2^2 - m_3^4 < 0 \, .
\ee
In addition, a certain number of conditions have to be satisfied
to have a stable minimum with the correct amount of symmetry
breaking and with unbroken colour, electric charge, baryon and lepton
number: for example, all the running squark and slepton masses have
to be
positive. In the whole process, a crucial role is played by the top
Yukawa coupling, which strongly influences the RGE for $m_2^2$.
For appropriate boundary conditions, the RGEs drive ${\cal B} < 0$ at
scales
$Q \sim G_F^{-1/2}$, whereas all the squark and slepton masses remain
positive
as desired, to give a phenomenologically acceptable breaking of the
electroweak
symmetry.

\subsection{Supersymmetric grand-unification}

The previous list has left out one of the most impressive arguments
in favour
of low-energy supersymmetry, i.e. the agreement of the generic
predictions of
supersymmetric grand unification with the extracted values of the
gauge
coupling constants at the electroweak scale. This topic has been
discussed at
great length by Zichichi$^{\ref{zichichi}}$ at this workshop, and I
will try to
give here my personal summary of the subject.

Starting from the boundary condition
\be
\label{gcu}
g_3 (M_U) =
g_2 (M_U) =
g_1 (M_U)
\equiv
g_U,
\ee
where $g_1 = \sqrt{5/3} \cdot g'$ as in most grand-unified models,
one can
solve the appropriate RGEs to obtain the running gauge coupling
constants $g_A
(Q)$ ($A=1,2,3$) at scales $Q << M_U$. At the one-loop level, and
assuming that
there are no new physics thresholds between $M_U$ and $Q$, one
finds$^{\ref{gqw}}$
\be
\label{running}
\frac{1}{g_A^2(Q)} = \frac{1}{g_U^2} +
\frac{b_A}{8 \pi^2} \log \frac{M_U}{Q}
\;\;\;\;\;
(A=1,2,3) \, ,
\ee
where the one-loop coefficients $b_A$  depend only on the $SU(3)_C
\times
SU(2)_L \times U(1)_Y$ quantum numbers of the light particle
spectrum. In
the MSSM
\be
\label{bsusy}
b_3 = - 3,
\;\;\;\;\;
b_2 = 1,
\;\;\;\;\;
b_1 = \frac{33}{5},
\ee
whereas in the SM
\be
\label{bsm}
b_3^0 = - 7,
\;\;\;\;\;
b_2^0 = - \frac{19}{6},
\;\;\;\;\;
b_1^0 =   \frac{41}{10}.
\ee
Starting from three input quantities at the electroweak scale, for
example
$\alpha_3 (m_Z)$, $\alpha_{em}^{-1} (m_Z)$ and $\sin^2 \theta_W
(m_Z)$, one can
perform consistency checks of the grand-unification hypothesis in
different
models.

In the minimal $SU(5)$ model$^{\ref{gg}}$, and indeed in any other
model
where Eq.~(\ref{gcu}) holds and the light-particle content is just
that
of the SM (with no intermediate mass scales between $m_Z$ and $M_U$),
Eqs.~(\ref{running}) and (\ref{bsm}) are incompatible with
experimental data.
This was first realized by noticing that the prediction $M_U \simeq
10^{14-15}
\gev$ is incompatible with the experimental data on nucleon decay.
Subsequently, also the prediction $\sin^2 \theta_W \simeq 0.21$ was
shown to
be in conflict with the experimental data, and this conflict became
even more significant after the recent LEP precision measurements.

In the MSSM, assuming for simplicity that all supersymmetric
particles
have masses of order $m_Z$, one obtains$^{\ref{drw}}$ $M_U \simeq
10^{16}
\gev$ (which increases the proton lifetime for gauge-boson-mediated
processes beyond the present experimental limits) and $\sin^2
\theta_W \simeq 0.23$. At the time of Ref.~\ref{drw}, when data were
pointing towards a significantly smaller value of $\sin^2 \theta_W$,
this
was considered by some a potential phenomenological shortcoming of
the MSSM.
The high degree of compatibility between data and supersymmetric
grand
unification became manifest only later$^{\ref{costa}}$, after
improved data on
neutrino-nucleon deep inelastic scattering were obtained; it was
recently
re-emphasized$^{\ref{grz},\ref{zichichi}}$ after the LEP precision
measurements. One should not forget, however, that unification of the
MSSM is
not the only solution that can fit the present extracted values of
the gauge
coupling constants at $Q=m_Z$: for example, non-supersymmetric models
with {\em
ad hoc} light exotic particles or intermediate symmetry-breaking
scales
could also do the job. The MSSM, however, stands out as the simplest
physically motivated solution.

If one wants to make the comparison between low-energy data and
the predictions of specific grand-unified models more precise,
there are several factors that should be further taken into account.
After the inclusion of higher-loop corrections and threshold effects,
Eq.~(\ref{running}) is (schematically) modified as follows
\be
\label{jpd}
\frac{1}{g_A^2(Q)} = \frac{1}{g_U^2} +
\frac{b_A}{8 \pi^2} \log \frac{M_U}{Q}+
\Delta_A^{th} + \Delta_A^{l>1}
\;\;\;\;\;
(A=1,2,3) \, .
\ee
In Eq.~(\ref{jpd}), $\Delta_A^{th}$ represents the so-called {\em
threshold
effects}, which arise whenever the RGEs are integrated across a
particle
threshold, and $\Delta_A^{l>1}$ represents the corrections due
to two- and higher-loop contributions to the RGEs. Both
$\Delta_A^{th}$
and $\Delta_A^{l>1}$ are scheme-dependent, so that one should be
careful
to compare data and predictions within the same renormalization
scheme.
The $\Delta_A^{th}$ receives contributions both from thresholds
around
the electroweak scale (top quark, Higgs boson, and in SUSY-GUTs also
the
additional particles of the MSSM spectrum) and from thresholds around
the
grand-unification scale (superheavy gauge and Higgs bosons, and in
SUSY-GUTs
also their superpartners). Needless to say, these last threshold
effects
can be computed only in the framework of a specific grand-unified
model,
and typically depend on a number of free parameters. Besides the
effects of
gauge couplings, $\Delta_A^{l>1}$ must include also the effects
of Yukawa couplings, since, even in the simplest mass-independent
renormalization schemes, gauge and Yukawa couplings mix beyond the
one-loop order. In minimal $SU(5)$ grand unification,
and for sensible values of the top and Higgs masses,
all these corrections are small and do not substantially
affect the conclusions derived from the na\"{\i}ve one-loop analysis.
This is no longer the case, however, for supersymmetric grand
unification.
First of all, one should notice that the MSSM by itself does not
uniquely
define a SUSY-GUT, whereas threshold effects and even the proton
lifetime
(owing to a new class of diagrams$^{\ref{proton}}$, which can be
originated in
SUSY-GUTs) become strongly model-dependent. Furthermore, the simplest
SUSY-GUT$^{\ref{dg}}$, containing only chiral Higgs superfields in
the $24$,
$5$ and $\ov{5}$ representations of $SU(5)$, has a severe problem in
accounting for the huge mass splitting between the $SU(2)$ doublets
and
the $SU(3)$ triplets sitting together in the $5$ and $\ov{5}$ Higgs
supermultiplets, and cannot reproduce correctly the observed pattern
of fermion masses and mixing angles. Threshold effects
are typically larger than in ordinary GUTs, because of the much
larger number
of particles in the spectrum, and in any given model they depend on
several
unknown parameters. Also two-loop effects of Yukawa couplings can be
quantitatively important in SUSY-GUTs, since they depend
not only on the top-quark mass, but also on the ratio $\tb = v_2/v_1$
of
the VEVs of the two neutral Higgs fields:
these effects become large for $m_t \simgt 140 \gev$ and $\tb \sim
1$, which
correspond to a strongly interacting top Yukawa coupling. All these
effects,
and others, have been recently re-evaluated$^{\ref{refined}}$.
The conclusion is that, even imagining a further
reduction in the experimental errors on the three gauge couplings,
it is impossible to claim indirect evidence for supersymmetry and
to predict the MSSM spectrum with any significant accuracy.  The only
safe statement is that, at the level of precision corresponding
to the na\"{\i}ve one-loop approximation, there is a remarkable
consistency
between experimental data and the prediction of supersymmetric
grand unification, with the MSSM R-odd particles roughly at the
electroweak scale. These conclusions are summarized in Fig.~2,
borrowed from
Ref.~\ref{barbieri}, which compares post-LEP and pre-LEP
uncertainties, both
theoretical and experimental, in the determination of $\sin^2
\theta_W (m_Z)$.
\begin{figure}
\vspace{16.0cm}
\caption{Comparison of theory and experiment in the determination of
the
electroweak mixing angle from the unification hypothesis, now and
before LEP
(from Ref.~22).}
\end{figure}

At this point it is worth mentioning how the unification constraints
can be
applied to the low-energy effective theories of four-dimensional
heterotic
string models. The basic fact to be realized is that the only free
parameter of
these models is the string tension, which fixes the unit of measure
of the
massive string excitations. All the other scales and parameters are
related to
VEVs of scalar fields, the so-called ${\em moduli}$, corresponding to
flat
directions of the scalar potential. In particular, there is a
relation between
the string mass $M_{\rm S} \sim \alpha'^{-1/2}$, the Planck mass
$M_{\rm P}
\sim  G_{\rm N}^{-1/2}$, and the unified coupling constant $g_U$,
which
reflects unification with gravity and implies that in any given
string vacuum
one has one more prediction than in ordinary field-theoretical grand
unification. In a large class of string models, one can write down an
equation
of the same form as (\ref{jpd}), and compute $g_U$, $M_U$,
$\Delta_A^{th}$,
$\ldots$ in terms of the relevant VEVs$^{\ref{strings}}$. In the
$\ov{DR}$
scheme one finds $M_U \simeq 0.5 \times g_U \times 10^{18} \gev$,
more than one
order of magnitude higher than the na\"{\i}ve extrapolations from
low-energy
data illustrated before. This means that significant threshold
effects are
needed in order to reconcile string unification with low-energy data:
for example, the minimal version of the flipped-$SU(5)$
model$^{\ref{flipped}}$
is by now ruled out$^{\ref{flopped}}$. To get agreement, one needs
some more
structure in the spectrum, either at the compactification scale or
in the form of light exotics, but even in this case one suffers
a loss of predictivity. However, unification constraints now stand as
a very important phenomenological test for any realistic string
model.

To conclude the discussion of supersymmetric grand unification, a few
more
words on proton decay seem appropriate. We have already stressed that
the RGE
of the MSSM imply a unification scale $M_U \simeq 10^{16} \gev$. This
means
that proton decay mediated by heavy vector bosons, which favours
decay modes
such as $p \to e^+ \pi^0$ and whose rate is proportional to
$M_U^{-4}$, is
suppressed to unobservable levels. On the other hand, an entirely new
possibility$^{\ref{proton}}$ arises in SUSY-GUTs. With the superfield
content of the MSSM, one can construct supersymmetric gauge-invariant
operators
with $\Delta B = \Delta L = \pm 1$ and mass dimension $d=5$. These
operators
can be generated, for example, by the exchange of some heavy Higgs
superfields
of minimal supersymmetric $SU(5)$. The nucleon decay amplitudes are
obtained by
dressing these operators by SUSY particle exchanges, to convert
sfermion lines
into light fermion lines. The resulting decay rate scales now as
$M_U^{-2}
M_{\rm SUSY}^{-2}$, with the actual numerical value depending both on
the
details on the SUSY-GUT model and on the details of the low-energy
sparticle
spectrum. In view of the first class of ambiguities, to my mind it is
not
particularly interesting to take minimal SUSY $SU(5)$ and to look for
constraints on the soft breaking terms from the limits on proton
decay.
On the other hand, an important generic feature emerges from the
symmetry
structure of the dimension-five operators: the dominant nucleon decay
modes
should involve strange particles in the final state, as in $p \to K^+
\ov{\nu}_{\mu}$. Detection of nucleon decay in one of these channnels
would
certainly be a very strong argument in favour of supersymmetric grand
unification.

\section{Supersymmetry searches}

\subsection{The particle spectrum of the MSSM}

In the $R$-even sector, the only new feature of the MSSM with respect
to
the SM is its extended Higgs sector, with two independent VEVs, $v_1
\equiv
\langle H_1^0 \rangle $ and $v_2 \equiv \langle H_2^0 \rangle $,
which can be
taken to be real and positive without loss of generality. Quarks of
charge
$2/3$ have masses proportional to $v_2$, quarks of
charge $1/3$ and charged leptons have masses proportional to $v_1$.
The W and Z masses are proportional to $\sqrt{v_1^2+v_2^2}$, which is
therefore fixed by their measured values. The remaining freedom is
conveniently parametrized by $\tan \beta \equiv v_2/v_1$, whose
allowed
range of variation in the MSSM is $1 \simlt \tan \beta \simlt
m_t/m_b$.
The physical states of the MSSM Higgs sector are three neutral bosons
(two CP-even, $h$ and $H$, and one CP-odd, $A$) and a charged boson,
$H^{\pm}$. It is important to realize that, at the tree level, all
Higgs
masses and couplings can be expressed in terms of two parameters
only:
a convenient choice is, for example, ($m_A, \tan \beta$). Radiative
corrections
to Higgs masses and couplings, however, can be large, as we shall
review later,
and have to be taken into account in phenomenological analyses.

In the $R$-odd sector of the MSSM, the spin-0 fields are the squarks
and
the sleptons. Neglecting intergenerational mixing, and leaving aside
the stop squarks for the moment, their masses can be easily
calculated
in terms of the fundamental parameters $m_0$, $m_{1/2}$ and $\tan
\beta$:
\begin{equation}
m_{\tilde{f}}^2 = m_f^2 + \tilde{m}^2_f  + m_D^2 ( \tilde{f} ),
\end{equation}
\begin{equation}
\tilde{m}^2_f = m_0^2 + C ( \tilde{f} ) m_{1/2}^2,
\end{equation}
\begin{equation}
m_D^2 ( \tilde{f} ) = m_Z^2 \cos 2 \beta ( T_{3L}^f - \sin^2 \theta_W
Q^f ),
\end{equation}
where, omitting generation indices, $f = [ q \equiv (u,u^c,d,d^c), l
\equiv (\nu,e), e^c]$ and $C(\tilde{q}) \sim 5-8$, $C(\tilde{l})
\simeq 0.5$, $C(\tilde{e^c}) \simeq 0.15$.

Among the spin-$1/2$ particles one finds the strongly interacting
gluinos,
$\tilde{g}$, whose mass is directly related to the $SU(2)$ and $U(1)$
gaugino
masses and to $m_{1/2}$ by
\begin{equation}
\label{gauginos}
{m_{\tilde{g}} \over \alpha_S}
\simeq
{M_2 \over \alpha_2}
\simeq
{M_1 \over \alpha_1}
\simeq
{m_{1/2} \over {\alpha_U}}
 \, ,
\end{equation}
where $\alpha_1 \equiv (5/3) g'^{\, 2} / (4 \pi)$ and $\alpha_U$ is
the gauge
coupling strength at the grand unification scale. The weakly
interacting
spin-$1/2$ particles are the $SU(2) \times U(1)$ gauginos
($\tilde{W}^{\pm}$;
$\tilde{W}^0$, $\tilde{B}$) and the higgsinos ($\tilde{H}^{\pm}$;
$\tilde{H}_1^0$, $\tilde{H}_2^0$). These interaction eigenstates
mix non-trivially via their mass matrices: the two charged mass
eigenstates, called {\it charginos}, are denoted by
$\tilde{\chi}^{\pm}_i$
($i=1,2$), and the four neutral mass eigenstates, called {\it
neutralinos},
by $\tilde{\chi}_k^0$ ($k=1,2,3,4$). All masses and couplings in the
chargino-neutralino sector can be described in terms of the three
parameters $m_{1/2}$ [which determines the $SU(2) \times U(1)$
gaugino
masses via eq.~(\ref{gauginos})], $\mu$ (the supersymmetric
Higgs-Higgsino mass term) and $\tan \beta$.
It should be noted that $\tilde{\chi}_1^0$, often denoted simply as
$\tilde
{\chi}$, is the favourite candidate for being the LSP. An alternative
candidate is $\tilde{\nu}_{\tau}$, but this is actually the LSP for a
much
smaller range of parameter space. Notice also that there is no
particular
reason to assume that $\tilde{\chi}$ is a pure photino, as is often
done in
phenomenological analyses.

In summary, the particle spectrum of the MSSM can be approximately
described in terms of five basic parameters:
\begin{itemize}
\item
The mass $m_A$ of the CP-odd neutral Higgs boson (or any other SUSY
Higgs mass)
\item
The ratio of VEVs $\tan \beta \equiv v_2 / v_1$
\item
The universal gaugino mass $m_{1/2}$, or equivalently
the gluino mass $m_{\tilde{g}}$
\item
The universal scalar mass $m_0$
\item
The supersymmetric Higgs-Higgsino mass $\mu$
\end{itemize}
Of course, the top-quark mass is undetermined, as in the SM. Also, as
we shall
see later, more subtleties have to be introduced for the description
of the
stop system.

\subsection{Searches for SUSY Higgs bosons}

We have already mentioned the fact that, at the classical level, the
Higgs
sector of the MSSM is very tightly constrained. However, as
extensively
discussed at this
workshop$^{\ref{haber},\ref{hempfling},\ref{ridolfi}}$, Higgs
boson masses and couplings are subject to large, finite radiative
corrections,
dominated by loops involving the top quark and its supersymmetric
partners.

To illustrate the case with a simple example, we can assume a
universal soft
SUSY-breaking squark mass, $\msq$, and negligible mixing in the stop
mass
matrix. The leading correction to the neutral CP-even mass matrix is
then
\be
\label{cpeven1}
\left( \Delta {\cal M}_R^2 \right)_{22}
=
{3 \over 8 \pi^2} {g^2 m_t^4 \over m_W^2 \sin^2 \beta} \log
\left( 1 + {\msq^2 \over m_t^2} \right) \, .
\ee
The most striking fact in Eq.~(\ref{cpeven1}) is that the correction
$( \Delta
{\cal M}_R^2 )_{22}$ is proportional to $(m_t^4/m_W^2)$ for fixed
$(\msq/m_t)$.
This implies that, for $m_t$ in the presently allowed range, the
tree-level
predictions for $m_h$ and $m_H$ can be badly violated, as for the
related
inequalities. The other free parameter in Eq.~(\ref{cpeven1}) is
$\msq$, but
the dependence on it is much milder. The result of
Eq.~(\ref{cpeven1}) has been
generalized to arbitrary values of the parameters in the stop mass
matrix, and
the effects of other virtual particles in the loops have been
included.
Renormalization-group methods have been used to resum the large
logarithms that
arise when the typical scale of supersymmetric particle masses,
$M_{\rm SUSY}$,
is much larger than $m_Z$. Two-loop corrections have been computed in
the
leading logarithmic approximation, and found to be small. After all
these
refinements, Eq.~(\ref{cpeven1}) still gives the most important mass
correction
in the most plausible region of parameter space.

The computation of radiative corrections has been extended to the
other
parameters of the MSSM Higgs sector. One-loop corrections to the
charged
Higgs mass have been computed and found to be small, at most a few
GeV,
for generic values of the parameters. In the case of Higgs boson
self-couplings, which control decays such as $H \to hh$, $H \to AA$
and $h \to
AA$ when they are kinematically allowed, radiative corrections can be
numerically large, being formally proportional to $(m_t/m_W)^4$.
Higgs
couplings to vector boson and fermions can be more easily handled: in
most
phenomenological studies, radiative corrections to these couplings
need to be
taken into account only approximately, by improving the tree-level
formulae
with one-loop-corrected values of the $H$--$h$ mixing angle,
$\alpha$, and with
running fermion masses, evaluated at the typical scale $Q$ of the
process under
consideration. Residual corrections have been computed and found to
be
numerically small in the experimentally interesting regions of
parameter
space.

We now move to the discussion of SUSY Higgs searches at present and
future
accelerators. For definiteness, when making numerical examples we
shall make
the same assumptions as for Eq.~(\ref{cpeven1}) and choose the
numerical values
$m_t = 140 \gev$, $\msq = 1 \tev$: for this parameter choice, the
maximum value of $m_h$, reached for $m_A \gg m_Z$ and $\tb \gg 1$, is
approximately 110 GeV, $O(20 \gev)$ larger than the tree-level upper
bound.
For given $m_A$ and $\tb$, the shift in $m_h$ can be as large as
$O(50 \gev)$,
when $\tb \sim 1$. In particular, after radiative corrections one can
have
not only $m_h > m_Z$, but also $m_h > m_A$.

As discussed by Clare$^{\ref{clare}}$ and Fisher$^{\ref{fisher}}$,
the relevant
processes for MSSM Higgs boson searches at LEP~I are $Z \to h Z^*$
and $Z \to h
A$, which play a complementary role, since their rates are
proportional to
$\sabsq$ and $\cabsq$, respectively. An important effect of radiative
corrections is to render possible, for some values of the parameters,
the decay
$h \to AA$, which would be kinematically forbidden according to
tree-level
formulae. Experimental limits that take radiative corrections into
account have
by now been obtained by the four LEP collaborations, using different
methods to
present and analyse the data, and different ranges of parameters in
the
evaluation of radiative corrections. An example is given in Fig.~3,
where the
cross-hatched area corresponds to the presently excluded region for
the
parameter choice $m_t = 140 \gev$, $\msq = 1 \tev$.
\begin{figure}
\vspace{8.0cm}
\caption{In the $(m_{h,H},m_A)$ plane, and for the parameter choice
$m_t = 140
\; {\rm GeV}$, $m_{\tilde{q}} = 1 \; {\rm TeV}$: the domain presently
excluded,
shown cross-hatched, and the domain which can be explored at LEP~II,
shown
hatched. The dash-dotted line is the kinematic limit for $HA$
associate
production at 500 GeV (from Ref.~31).}
\end{figure}

The situation in which the impact of radiative corrections is most
dramatic is the search for MSSM Higgs bosons at LEP~II, as discussed
at this
workshop by Katsanevas$^{\ref{katsanevas}}$. At the time when only
tree-level
formulae were available, there was hope that LEP could completely
test the MSSM
Higgs sector. According to tree-level formulae, in fact, there should
always be
a CP-even Higgs boson with mass smaller than $m_Z$ ($h$) or very
close to it
($H$), and significantly coupled to the $Z$ boson. However, this
result can
be completely upset by radiative corrections. A detailed evaluation
of the
LEP~II discovery potential can be made only if crucial theoretical
parameters
(such as the top-quark mass and the various soft
supersymmetry-breaking masses)
and experimental parameters (such as the centre-of-mass energy, the
luminosity,
and the $b$-tagging efficiency) are specified. Taking for example
$\sqrt{s} =
190 \gev$, and the parameter choice of Fig.~3, there is a region of
parameter
space where the associated production of a $Z$ and a CP-even Higgs
can be
pushed beyond the kinematical limit. Associated $hA$ production could
be a
useful complementary signal, but obviously only for $m_h+m_A<
\sqrt{s}$.
Associated $HA$ production is typically negligible at these energies.
The
hatched area of Fig.~3 shows the domain accessible to LEP~II for the
mentioned
parameter choice, for an integrated luminosity of $500 \pb^{-1}$ and
for a
detector similar to the ALEPH detector at LEP: one can see that the
theoretically allowed parameter space cannot be fully tested.

Of course, one should keep in mind that there is, at least in
principle, the
possibility of further extending the maximum LEP energy up to values
as high as
$\sqrt{s} \simeq 230$--$240 \gev$, at the price of introducing more
(and more
performing) superconducting cavities into the LEP tunnel. More
boldly, one
can consider the possibility of an $e^+ e^-$ linear collider with
$\sqrt{s}
\sim 500 \gev$ and a luminosity of order $10^{33} \, {\rm cm}^{-2}
{\rm
sec}^{-1}$: a detailed study of the discovery potential of such a
collider has
been presented at this workshop by Grivaz$^{\ref{grivaz}}$. Among the
relevant
production mechanisms there are those already mentioned for LEP~II:
(a)~$\epem
\to h Z$, (b)~$\epem \to H Z$, (c)~$\epem \to h A$, (d)~$\epem \to H
A$; in
addition, one can consider $WW$ and $ZZ$ fusion: (e)~$\epem \to h \nu
\ov{\nu}
\; {\rm or} \; h \epem$, (f)~$\epem \to H \nu \ov{\nu} \; {\rm or} \;
H \epem$.
 Considering the domain that will remain unexplored if the
centre-of-mass
energy at LEP~II is limited to 190 GeV, there are four main
configurations (see
Fig.~3): in (1) $h$ is SM-like and accessible via (a) and (e); in (2)
one has
in addition the possibility of detecting (d); in (3) the observable
processes
are (b), (c) and (f); in (4) all processes are kinematically allowed
and only
moderately suppressed with respect to the SM case. Also, in regions
(2), (3)
and (4) one can observe pair production of charged Higgses: (g)
$\epem \to H^+
H^-$. In summary, such a linear $\epem$ collider would allow for a
complete
exploration of the MSSM parameter space: if the MSSM is indeed
correct, one
could expect the guaranteed detection of at least one neutral Higgs
state and
the concrete possibility of a detailed spectroscopy of the Higgs
sector.

Another interesting possibility offered by a linear $\epem$ collider
is the
study of $\gamma \gamma$ collisions at very high energy and
luminosity, thanks
to a back-scattered laser beam facility. The physics impact of such a
machine
on the SUSY-Higgs sector has been discussed by
Gunion$^{\ref{gunion}}$ at this
meeting, who emphasized its complementarity with respect to the
$\epem$ mode.

The next question, which was discussed by Kunszt$^{\ref{kunszt}}$ and
also by
Gunion$^{\ref{gunion}}$, is whether the LHC/SSC can explore the full
parameter
space of the MSSM Higgs bosons. The analysis is complicated by the
fact that
the $R$-odd particles could play a role both in the production (via
loop
diagrams) and in the decay (via loop diagrams and as final states) of
the MSSM
Higgs bosons. For simplicity, one can concentrate on the most
conservative case
in which all $R$-odd particles are heavy enough not to play any
significant
role. Still, one has to perform a separate analysis for each
$(m_A,\tb)$ point,
to include radiative corrections (depending on additional parameters
such as
$m_t$ and $\msq$), and to consider Higgs boson decays involving other
Higgs
bosons.

\begin{figure}
\vspace{9.0cm}
\caption{Pictorial representation of the LHC/SSC discovery potential
in the
$(m_A, \tan \beta)$ plane, characterizing the Higgs sector of the
MSSM, for the
parameter choice described in the text (from Ref.~33).}
\end{figure}

The most promising signals at the LHC/SSC are  $h,H \to \gamma
\gamma$
(inclusive or in association with a $W$ boson or with a $t \ov{t}$
pair, giving an extra isolated lepton in the final state), $H \to Z Z
\to 4
l^{\pm}$, $A,H \to \tau^+ \tau^-$ and $t \to b H^+ \to b \tau^+
\nu_{\tau}$.
A pictorial representation of the LHC/SSC discovery potential
corresponding to
the different processes is given in Fig.~4, which also shows, as
dashed lines,
contours associated with a `pessimistic' and an `optimistic' estimate
of the
LEP~II sensitivity. In summary, a global look at Fig.~4 shows that
there is a
high degree of complementarity between the regions of parameter space
accessible to LEP~II and to the LHC/SSC. However, for our
representative choice
of parameters, there is a non-negligible region of the $(m_A,\tb)$
plane that
is presumably beyond the reach of LEP~II and of the LHC/SSC. This
potential
problem could be solved, as we said before, by a further increase of
the LEP~II
energy beyond the reference value of $\sqrt{s} \simlt 190 \gev$ or by
a
high-energy linear $\epem$ collider. One should also keep in mind
that
indirect information on the particle spectrum of the MSSM, including
its
extended Higgs sector, could come from lower-energy precision data.
An
interesting effect, mentioned at this workshop by
Haber$^{\ref{haber}}$, is the
contribution of the charged-Higgs loop to the rare decay $b \to s
\gamma$,
which in the SM proceeds via a $W$-boson loop. The theoretical and
experimental
errors on the inclusive radiative $B$-decay could already be small
enough to
put non-trivial constraints on the particle spectrum of the MSSM. In
particular, in the limit of very heavy $R$-odd particles one could
identify an
excluded region in the $\matb$ plane, corresponding to low values of
$m_{H^\pm}$: the precise form of such a region strongly depends on
the assumed
theoretical uncertainties.

One should also mention that a complete study of the SUSY-Higgs
phenomenology
cannot neglect the possibility of a relatively light spectrum of
$R$-odd
particles: as reported by Baer and Tata$^{\ref{baertata}}$, in this
case one
expects a worsening of the standard signals, which could be
compensated,
however, by the appearance of new signals related to Higgs decays
into pairs of
$R$-odd particles.

\subsection{Searches for $R$-odd particles}

The phenomenology of $R$-odd SUSY particles has been discussed in
many
contributions to this workshop$^{\ref{kobayashi}, \ref{clare},
\ref{fisher} -
\ref{grivaz}, \ref{baertata} - \ref{orava}}$.

The present status of experimental searches for supersymmetry is a
collection
of negative results, which translate into limits on the MSSM
parameters.

LEP experiments have searched$^{\ref{fisher},\ref{clare}}$ for a
variety of
possible supersymmetric $Z$ decays
\be
Z \longrightarrow \tilde{l}^{\pm} \tilde{l}^{\mp} \, , \; \tilde{\nu}
\,
\tilde{\ov{\nu}} \, , \; \tilde{\chi}^{\pm} \tilde{\chi}^{\mp} \, ,
\;
\tilde{\chi}^0_i \tilde{\chi}^0_j \, , \; \tilde{q} \, \tilde{\ov{q}}
\, ,
\ee
both directly and indirectly (via measurements of the $Z$ line
shape).
A crude summary is that $R$-odd particles weighing much less than
$m_Z/2$ are
in general excluded, with some possible exceptions, corresponding to
particles
with strongly suppressed couplings to the $Z$. The first exception
are the
lightest neutralinos  $\tilde{\chi}_1^0$ and $\tilde{\chi}_2^0$: for
small
values of $m_{1/2}$ and $\tb$, one can have the charginos and the
heavier
neutralinos beyond the kinematical limit of LEP~I, and the two
lightest
neutralinos with dominant gaugino components and therefore decoupled
from the
$Z$ boson. The LEP lower bounds on the masses of $\tilde{\chi}_1^0$
and
$\tilde{\chi}_2^0$ are of the order of 20 and 45 GeV, respectively,
for large
values of $\tb$, but evaporate for values of $\tb$ sufficiently close
to unity.
The second exception is the lighter stop
state$^{\ref{kobayashi},\ref{fisher},\ref{hidaka}}$. In the stop mass
matrix,
the mixing term
between $\tilde{t}_L$ and $\tilde{t}_R$ is proportional to the
top-quark mass,
so it cannot always be neglected. It is quite possible that the
lighter stop
state, $\tilde{t}_1 \equiv \cos \theta_t \, \tilde{t}_L + \sin
\theta_t \,
\tilde{t}_R$, is significantly lighter than the top quark and the
remaining
squarks. In this case, its coupling to the $Z$ boson is proportional
to
$[\cos^2 \theta_t - (4/3) \sin^2 \theta_W]$, so that for certain
values of the
mixing angle one can still have $m_{\tilde{t}_1} \simeq 25 \gev$. The
third and
final exception is the gluino$^{\ref{kunszt}}$, which does not have
any
tree-level coupling to the $Z$ boson: no limit on the gluino mass has
been
extracted from LEP data so far.

\begin{figure}
\vspace{8.0cm}
\caption{CDF limits on squarks and gluinos assuming no cascade
decays.}
\end{figure}

In the case of squarks and gluinos, additional information comes from
the
experiments at $p \ov{p}$ colliders. In this case, limits are
considerably
more model-dependent than at LEP, due to the complicated pattern of
cascade
decays that can arise$^{\ref{baertata},\ref{bartl}-\ref{barnett}}$.
Figure~5
shows the squark and gluino mass limits assuming a light LSP
($m_{\tilde{\chi}}
< 15 {\rm \; GeV}$), six mass-degenerate squarks, and no cascade
decays.
However, once the MSSM squark and gluino branching ratios are
introduced into
the analysis, the above limits can be considerably degraded, as can
be seen in
the example of Fig.~6, corresponding to a representative parameter
choice in
the
chargino-neutralino sector. One should also keep in mind that  $p
\ov{p}$-collider searches are not sensitive to very light gluino
masses, which
must be excluded by different methods. At present, it seems very
difficult to
exclude gluinos weighing between 3 and 4~GeV with lifetimes around
$10^{-13}$
s, and gluinos weighing 3~GeV or more and having lifetimes between
about
$10^{-8}$ and $10^{-10}$~s.

\begin{figure}
\vspace{8.0cm}
\caption{CDF limits on squarks and gluinos with the effects of
cascade decays
taken into account, for the parameter choice $\mu = - 250$ GeV, $\tan
\beta =
2$ and $m_H= 500$ GeV.}
\end{figure}

Future accelerators should allow for great progress in the search for
$R$-odd
supersymmetric particles. At LEP~II, one should be sensitive to pair
production
of sleptons and charginos almost up to the kinematical
threshold$^{\ref{katsanevas},\ref{grivaz}}$. At the LHC and the SSC,
one should
be able to search for squarks and gluinos up to masses of the order
of 1~TeV or
more$^{\ref{baertata},\ref{barnett}}$. This should allow the coverage
of most
of the theoretically plausible region of parameter space.

\section{Theoretical outlook and conclusions}

Among the open theoretical problems, the most important and
challenging one is
to understand the mechanism of spontaneous supersymmetry breaking and
the
origin of the hierarchy. These problems cannot be addressed, by
definition,
within the MSSM, where supersymmetry is broken explicitly and the
soft breaking
terms are controlled by arbitrary input parameters. Present
theoretical
ideas$^{\ref{kounnas}}$ and phenomenological requirements favour the
possibility that supersymmetry is spontaneously broken in the {\em
hidden
sector} of some underlying supergravity (or superstring) model,
communicating
with the {\em observable} sector (the one containing the states of
the MSSM)
only via gravitational interactions. As for the precise mechanism of
spontaneous breaking of local supersymmetry, there are several
suggestions,
among which non-perturbative phenomena (such as gaugino condensation)
and
string constructions (such as coordinate-dependent
compactifications), but
none of them has yet reached a fully satisfactory formulation. If one
day a
convincing mechanism will be found, by taking the low-energy limit it
will be possible to predict the mass parameters of the MSSM, with
enormous
enhancement in predictive power.

Leaving aside this open theoretical problem, one can say that
low-energy
supersymmetry, incarnated in the MSSM, stands out as a theoretically
motivated,
phenomenologically acceptable and calculationally well-defined
extension of the SM. To the eyes of many observers, its plausibility
has
increased over the years. On the one hand, experimental searches have
excluded
until now only a relatively small part of its natural parameter
space. On the
other hand, electroweak precision measurements nicely fit an
elementary Higgs
sector and supersymmetric grand unification.

The phenomenological properties of the MSSM are by now well
understood.
Accurate computations of cross-sections and branching ratios for the
MSSM
particles are available, as functions of the model parameters. A lot
of
simulation work has been and is being performed for present and
future
colliders. At the level of the MSSM, the present challenge is mainly
an
experimental one. LEP~I, which has impressively improved the previous
limits
on the weakly interacting MSSM particles, has almost saturated its
discovery
potential. The next big step will occur at LEP~II, which should allow
the
search for charginos and charged sleptons up to masses of order
80--90 GeV,
and significant progress in the search for supersymmetric Higgs
bosons.
The present and future runs of the Tevatron collider should
significantly
push the sensitivity to squarks and gluinos, up to masses of order
200--300
GeV. If the idea of low-energy supersymmetry is correct, a discovery
is not
unlikely already at this stage. A negative result, however, would not
yet
discourage the SUSY advocates. Only the LHC and the SSC, and perhaps
a linear
$e^+ e^-$ collider with $\sqrt{s} \simgt 500 \gev$, will be able to
perform a
decisive test: the former should be able to probe squark and gluino
masses
up to 1--2 TeV, the latter should be sensitive to sleptons and
charginos
up to 200 GeV and more, and to SUSY Higgs bosons in any plausible
model. A
positive signal would open up an exciting new generation of
experiments, a
negative one would presumably push low-energy supersymmetry into
oblivion.

\section*{Acknowledgements}
I would like to thank Prof.~A.~Zichichi for the warm hospitality in
Erice,
the Directors of the Workshop, Prof.~L.~Cifarelli and Prof.~V.~Khoze,
for their
successful scientific organization, and Dr.~F.~Anselmo for his
helpful
assistance.

\newpage

\section*{References}
\begin{enumerate}
\item
\label{fayet}
P.~Fayet, contribution to these proceedings, and references therein.
\item
\label{kounnas}
C.~Kounnas, contribution to these proceedings, and references therein
\item
\label{dreiner}
H.~Dreiner, contribution to these proceedings, and references
therein.
\item
\label{kobayashi}
T.~Kobayashi, contribution to these proceedings, and references
therein.
\item
\label{kane}
G.L.~Kane, contribution to these proceedings, and references therein.
\item
\label{ibanez}
L.E.~Ib\'a\~nez, contribution to these proceedings, and references
therein.
\item
\label{antoniadis}
I.~Antoniadis, contribution to these proceedings, and references
therein.
\item
\label{nonuniv}
L.E.~Ib\'a\~nez and D.~L\"ust, Nucl. Phys. B382 (1992) 305;
V.S.~Kaplunovsky
and J.~Louis, Phys. Lett. B306 (1993) 269; B.~de Carlos, J.A.~Casas
and C.~Mu\~noz, preprint CERN-TH.6681/92.
\item
\label{haber}
H.E.~Haber, contribution to these proceedings, and references
therein.
\item
\label{wagner}
C.E.M.~Wagner, contribution to these proceedings, and references
therein.
\item
\label{roszkowski}
L.~Roszkowski, contribution to these proceedings, and references
therein.
\item
\label{lahanas}
A.B.~Lahanas, contribution to these proceedings, and references
therein.
\item
\label{zichichi}
A.~Zichichi, contribution to these proceedings, and references
therein.
\item
\label{gqw}
H. Georgi, H.R. Quinn and S. Weinberg, Phys. Rev. Lett. 33 (1974)
451.
\item
\label{gg}
H. Georgi and S.L. Glashow, Phys. Rev. Lett. 32 (1974) 438.
\item
\label{drw}
S.~Dimopoulos, S.~Raby and F.~Wilczek, Phys. Rev. D24 (1981) 1681;
L.E.~Ib\'a\~nez and G.G.~Ross, Phys. Lett. 105B (1981) 439.
\item
\label{costa}
U.~Amaldi, A.~B\"ohm, L.S.~Durkin, P.~Langacker, A.K.~Mann,
W.J.~Marciano,
A.~Sirlin and H.H.~Williams, Phys. Rev. D36 (1987) 1385; G.~Costa,
J.~Ellis,
G.L.~Fogli, D.V.~Nanopoulos and F.~Zwirner, Nucl. Phys. B297 (1988)
244.
\item
\label{grz}
G.~Gamberini, G.~Ridolfi and F.~Zwirner, Nucl. Phys. B331 (1990) 331;
J.~Ellis, S.~Kelley and D.V.~Nanopoulos, Phys. Lett. B249 (1990) 442
and B260 (1991) 131; P.~Langacker and M.~Luo, Phys. Rev. D44 (1991)
817;
U.~Amaldi, W.~de~Boer and H.~F\"urstenau, Phys. Lett. B260 (1991)
447.
\item
\label{proton}
S.~Weinberg, Phys. Rev. D26 (1982) 287;
N.~Sakai and T.~Yanagida, Nucl. Phys. B197 (1982) 533.
\item
\label{dg}
S.~Dimopoulos and H.~Georgi, Nucl. Phys. B193 (1981) 150;
N.~Sakai, Z.~Phys. C11 (1982) 153.
\item
\label{refined}
R. Barbieri and L.J. Hall, Phys. Rev. Lett. 68 (1992) 752;
J. Ellis, S. Kelley and D.V. Nanopoulos, Nucl. Phys. B373 (1992)
55;
G.G. Ross and R.G. Roberts, Nucl. Phys. B377 (1992) 571;
J.~Hisano, H.~Murayama and T.~Yanagida, Phys. Rev. Lett. 69 (1992)
1014;
R.~Arnowitt and P.~Nath, Phys. Rev. Lett. 69 (1992) 725 and Phys.
Lett. B287
(1992) 89;
K.~Hagiwara and Y.~Yamada, Phys. Rev. Lett. 70 (1993) 709;
L.J.~Hall and U.~Sarid, Phys. Rev. Lett. 70 (1993) 2673;
P.~Langacker and N.~Polonsky, Phys. Rev. D47 (1993) 4028.
\item
\label{barbieri}
R.~Barbieri, preprint CERN-TH.6863/93.
\item
\label{strings}
V.S.~Kaplunovsky, Nucl. Phys. B307 (1988) 145;
L.~Dixon, V.S.~Kaplunovsky and J.~Louis, Nucl. Phys. B355 (1991) 649;
J.P.~Derendinger, S.~Ferrara, C.~Kounnas and F.~Zwirner, Nucl. Phys.
B372
(1992) 145 and Phys. Lett. B271 (1991) 307;
G.~Lopez-Cardoso and B.A.~Ovrut, Nucl. Phys. B369 (1992) 351;
J.~Louis, preprint SLAC-PUB-5527 (1991);
I.~Antoniadis, K.S.~Narain and T.~Taylor,
Phys. Lett. B267 (1991) 37.
\item
\label{flipped}
I. Antoniadis, J. Ellis, J. Hagelin and D.V. Nanopoulos, Phys. Lett.
B231
(1989) 65, and references therein.
\item
\label{flopped}
I. Antoniadis, J. Ellis, R. Lacaze and D.V. Nanopoulos, Phys. Lett.
B268
(1991) 188.
\item
\label{hempfling}
R.~Hempfling, contribution to these proceedings, and references
therein.
\item
\label{ridolfi}
G.~Ridolfi, contribution to these proceedings, and references
therein.
\item
\label{clare}
R.~Clare, contribution to these proceedings, and references therein.
\item
\label{fisher}
P.~Fisher, contribution to these proceedings, and references therein.
\item
\label{katsanevas}
S.~Katsanevas, contribution to these proceedings, and references
therein.
\item
\label{grivaz}
J.-F.~Grivaz, contribution to these proceedings, and references
therein.
\item
\label{gunion}
J.F.~Gunion, contribution to these proceedings, and references
therein.
\item
\label{kunszt}
Z.~Kunszt, contribution to these proceedings, and references therein.
\item
\label{baertata}
H.~Baer and X.~Tata, contribution to these proceedings, and
references therein.
\item
\label{hidaka}
K.~Hidaka, contribution to these proceedings, and references therein.
\item
\label{bartl}
A.~Bartl, contribution to these proceedings, and references therein.
\item
\label{majerotto}
W.~Majerotto, contribution to these proceedings, and references
therein.
\item
\label{barnett}
R.M.~Barnett, contribution to these proceedings, and references
therein.
\item
\label{orava}
R.~Orava, contribution to these proceedings, and references therein.
\end{enumerate}
\end{document}